# ELUCIDATING THE METHYLAMMONIUM (MA) CONFORMATION IN $MAPbBr_3$ PEROVSKITE WITH APPLICATION IN SOLAR CELLS

*Carlos A. López,[a,b] María Victoria Martínez-Huerta,[c] María Consuelo Alvarez-Galván,[c] Paula Kayser,[a] Patricia Gant,[d] Andres Castellanos-Gomez,[a] María T. Fernández-Díaz,[e] Francois Fauth,[f] and José A. Alonso*[*,a]*

[a]Instituto de Ciencia de Materiales de Madrid, CSIC, Cantoblanco 28049 Madrid, Spain.

[b]INTEQUI, Universidad Nacional de San Luis, CONICET, Facultad de Química, Bioquímica y Farmacia, Chacabuco y Pedernera, 5700 San Luis, Argentine.

[c]Instituto de catálisis y Petroleoquímica, CSIC, Cantoblanco 28049 Madrid, Spain.

[d]Instituto Madrileño de Estudios Avanzados en Nanociencia (IMDEA Nanociencia), Campus de Cantoblanco, E-28049 Madrid, Spain.

[e]Institut Laue Langevin, BP 156X, Grenoble, F-38042, France.

[f]CELLS–ALBA synchrotron E-08290 Cerdanyola del Valles, Barcelona, Spain.

*Corresponding author: (ja.alonso@icmm.csic.es)

**KEYWORDS:** Hybrid perovskite, NPD, SXRPD, Photocurrent, Optoelectronics.

## ABSTRACT

Hybrid organic-inorganic perovskites, $MAPbX_3$ (X= halogen), containing methylammonium (MA: $CH_3$-$NH_3^+$) in the large voids conformed by the $PbX_6$ octahedral network, are the active absorption materials in the new generation of solar cells. $CH_3NH_3PbBr_3$ is a promising alternative with a large band-gap that gives rise to a high open circuit voltage. A deep knowledge of the crystal structure and, in particular, the MA conformation inside the perovskite cage across the phase transitions undergone below room temperature, seems essential to establish structure-property correlations that may drive to further improvements. The presence of protons requires the use of neutrons, combined with synchrotron XRD data that help to depict subtle symmetry changes undergone upon cooling. We present a consistent picture of the structural features of this fascinating material, in complement with photocurrent measurements from a photodetector device, demonstrating the potential of $MAPbBr_3$ in optoelectronics.

## INTRODUCTION

Organic–inorganic hybrid perovskites have emerged as promising materials for the next generation of solar cells because of their ease of fabrication and performances rivalling the best thin-film photovoltaic devices.[1-5] The introduction of $CH_3NH_3PbX_3$ (X = Br and I) by Miyasaka *et.al.* as a sensitizer in an electrolyte-based solar cell structure marked the beginning of perovskite-based photovoltaics.[2] However, the power conversion efficiency (PCE) and cell stability were poor due to the corrosion of the perovskites by the liquid electrolyte. A key advance was made in 2012 by replacing the liquid electrolyte with a solid hole transporting material, which resulted in both high PCE of 9.7% and enhanced cell stability.[3] Since then, an intensive research has been devoted to the improvement of halide perovskite-based solar cells, where a variety of cell architectures have been developed increasing the PCEs to 22.1% over the past years.[6]

Fundamental studies have revealed the superior optical and electrical properties of halide perovskites, including the tunability of the band-gap by varying the halide (i.e., bromide, iodide) composition of the perovskite precursor solution,[7-9] the high absorption coefficient[10] and the long lifetime of photogenerated species.[11] $CH_3NH_3PbI_3$-based perovskite solar cells have been a primary focus due to their near-complete visible light absorption in films <1 μm and their fast charge extraction rates.[11,12] However, the poor stability of $CH_3NH_3PbI_3$ and rapid degradation in humidity has remained a major obstacle for commercialization.[13,14] $CH_3NH_3PbBr_3$ is a promising alternative to $CH_3NH_3PbI_3$ with a large band-gap of 2.2 eV, which gives rise to a high open circuit voltage (Voc ≈ 1.2–1.5 V).[15,16] Their long exciton diffusion length (>1.2 μm) enables good charge transport in devices.[17] In addition, $CH_3NH_3PbBr_3$ demonstrates higher stability towards air and moisture due to its stable cubic phase and low ionic mobility relative to the pseudocubic $CH_3NH_3PbI_3$, in which inherent lattice strain provides an avenue for increased diffusion.[7,17-19] These features may compensate for a relatively large exciton binding energy (76 meV) and reduced light absorption beyond its band edge at 550 nm, accounting for a more limited efficiencies of $CH_3NH_3PbBr_3$ solar cells.[7,18,20-22]

In parallel with the evaluation of the performance contributed by a particular chemical substitution, it is necessary to undertake a suitable crystallographic characterization in the same state or conditions in which the sample will be used. It is of paramount importance to unveil the details of the crystal structure in relation to the physical behaviour, such as $CH_3$-$NH_3^+$ (methylammonium - MA) delocalisation, anisotropic displacement factors, tilting of polyhedra, etc. This knowledge is essential to establish relationships between the structures and the macroscopic phenomenology. $MAPbBr_3$ was previously studied by diffraction techniques in single crystal form by X ray or in deuterated sample by neutron beam.[23-25] In this work, we study the crystallographic features in a powdered, non-deuterated sample from neutron and synchrotron X ray diffraction at different temperatures. From these techniques, we describe the evolution of the orientation of MA group in the 120-295 K temperature range.

**EXPERIMENTAL SECTION**

$MAPbBr_3$ was crystallized as an orange microcrystalline powder from a solution of stoichiometric amounts of $PbBr_2$ and MABr in dimetilformamide. The crystals were ground prior to the diffraction experiments. Laboratory XRPD patterns were collected on a Brucker D5 diffractometer with KαCu (λ = 1.5418 Å) radiation; the 2θ range was 4° up to 90° with increments of 0.03°. The thermal evolution of the crystallographic structure was studied by synchrotron X-ray powder diffraction (SXRPD) at 120, 150, 180, 210, 240, 270 K and room temperature (298 K). SXRPD patterns were collected in high angular resolution mode (so-called MAD set-up) on the MSPD diffractometer in ALBA synchrotron at Barcelona, Spain, selecting an incident beam with 38 keV energy, λ= 0.3252 Å.[26] The sample was contained in a 0.3 mm diameter quartz capillary that was rotating during the data acquisition. Additionally, a NPD pattern at RT was collected on the D2B diffractometer, Laue Langevin Institut (ILL), Grenoble, France with a wavelength of 1.594 Å. The non-deuterated sample was contained in a 6 mm diameter vanadium cylinder. The coherent scattering lengths for Pb, Br, C, N and H were, 9.405, 6.795, 6.646, 9.36 and -3.739 fm, respectively. XRPD, SXRPD and NPD diffraction patterns were analyzed with the Rietveld method using the FullProf program.[27,28] A photodetector device was fabricated by drop-casting the perovskite solution in dimetilformamide onto Au/Cr pre-patterned electrodes with a gap of 10 µm, and drying in a hot plate at 100ºC.

## RESULTS AND DISCUSSION

Laboratory XRPD patterns at RT are similar to those previously reported for this phase;[23] no impurities were observed (Fig. S1 at the Supporting Information). The crystal structure was reported at RT as cubic in the space group $Pm\bar{3}m$ using x-ray single crystal diffraction.[23] Attempts to fit the pattern to this symmetry were not satisfactory, since the sample exhibits strong preferred orientation that cannot be simulated. However, a Le-Bail fit to the cubic symmetry was successful (Fig. S1), obtaining a cell parameter a =5.9595(1) Å.

Even though, the crystal structure was successfully analyzed from the SXRPD patterns: using high-energy synchrotron x-rays in transmission mode contributed to a better powder averaging from rotating capillaries, hence minimizing the preferred orientation. The thermal evolution of the crystallographic structure was followed between 120 K and RT. In this temperature range two phase transitions were observed, as previously reported by Swaison et.al.[23] The patterns collected at RT, 270 and 240 K were refined as cubic in $Pm\bar{3}m$ space group, at 210 and 180 K as tetragonal in *I*4/*mcm* space group, and at 120 K as orthorhombic in *Pnma* space group. At 150 K, transient diffraction lines were observed, which could not be indexed either as tetragonal or as orthorhombic. Figure 1 shows the thermal evolution of selected diffraction lines.

In the high temperature model ($Pm\bar{3}m$), the lead and bromine atoms were placed in 1*a* (0,0,0) and 3*d* (1/2,0,0) positions, respectively. Then, a difference Fourier synthesis map was obtained from the observed and calculated powder diffraction data. Figure 2.a shows the electron density as a contour surface at 2.2 e for the pattern collected at RT, corresponding to C and N atoms of the MA unit. The observed positive density suggests that C/N atoms are located at large multiplicity 24*i* (0.5,y,y) sites. This fact unveils that MA is delocalized between six different positions all aligned along [110] direction. After adding the C/N atoms to the structural model, the structure can be correctly refined at RT. The Rietveld plot is included in Fig. S2. When the same procedure is carried out at 270 and 240 K, several differences in the electron density were observed. Figure 2.b illustrates these differences in the zone where the MA unit is located. At 270 K the delocalization of MA is similar to RT (along [110] directions) but a clear decrease in the thermal vibrations is revealed. However, a drastic change in the delocalization is observed at 240 K: Figure 2.b shows that the MA units are aligned along [100] directions, reducing their delocalization as temperature decreases. The best structural fits, taking into account the MA delocalization, are displayed in Figure S2, and the corresponding crystallographic data are listed in Tables S1, S2 and S3 (Supplementary information), for RT, 270 and 240 K, respectively.

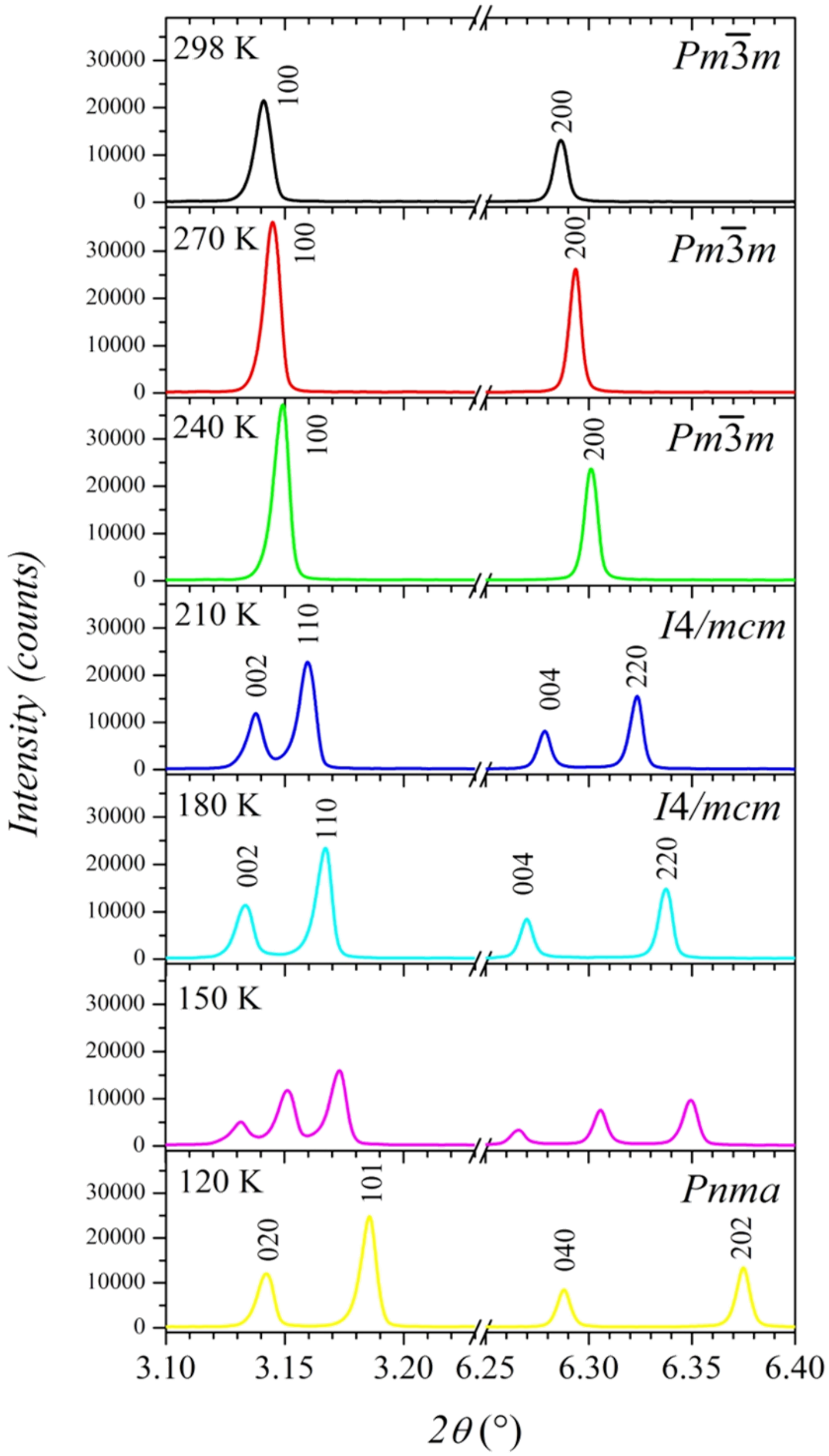

**Figure 1.** Thermal evolution of selected diffractions lines in which the phase transitions are evidenced, from SXRPD data.

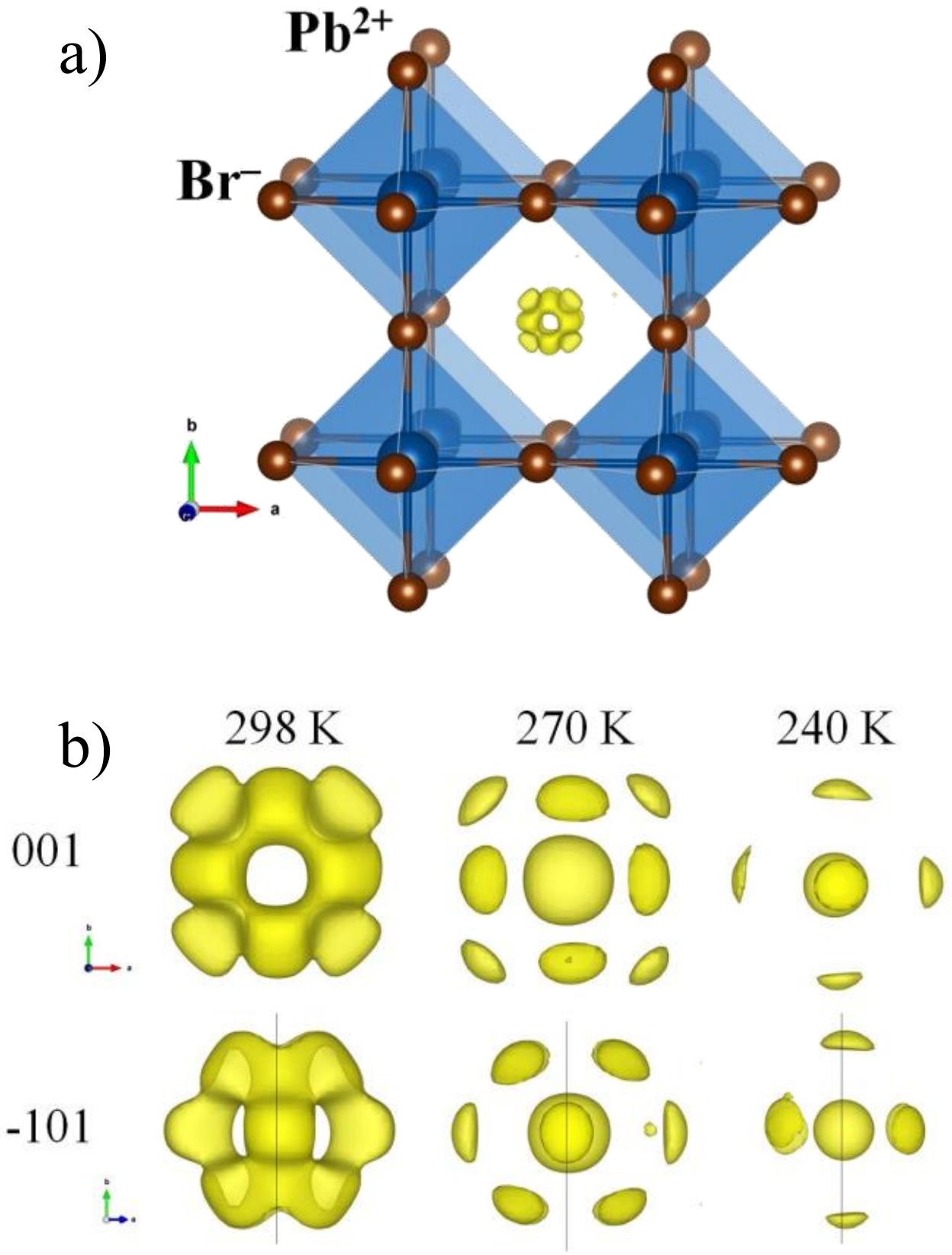

**Figure 2.** a) Schematic view of lead (blue spheres) and bromine (brown spheres) atoms in the cubic system. The isosurface of electron density at 2.2 e is shown in yellow. b) Thermal evolution of electron density (in the three cases at 2.2 e) along 001 and -101 directions.

The patterns collected at 210 and 180 K show a clear tetragonal distortion, see Figure 1, and the structure was refined (*I*4/*mcm* space group) placing Pb at 4*c* (0,0,0) and Br at 4*a* (0,0,1/4) and 8*h* (x,x+1/2,0) positions. At this point, the difference Fourier synthesis maps at both temperatures unveiling the missing electron density indicate that the MA units are delocalized as in the cubic system (see Figure 3). Two different densities are determined, corresponding to 16*l* Wyckoff site. Therefore, C and N atoms were located at this position and fitted with a distance constraint of 1.49 Å (standard deviation of 0.04 Å) in order to keep up the chemical restraints of MA. The structures at 210 and 180 K were correctly refined; the Rietveld plots are displayed in Figure S3. The crystallographic parameters at 210 and 180 K are listed in Tables S4 and S5, respectively.

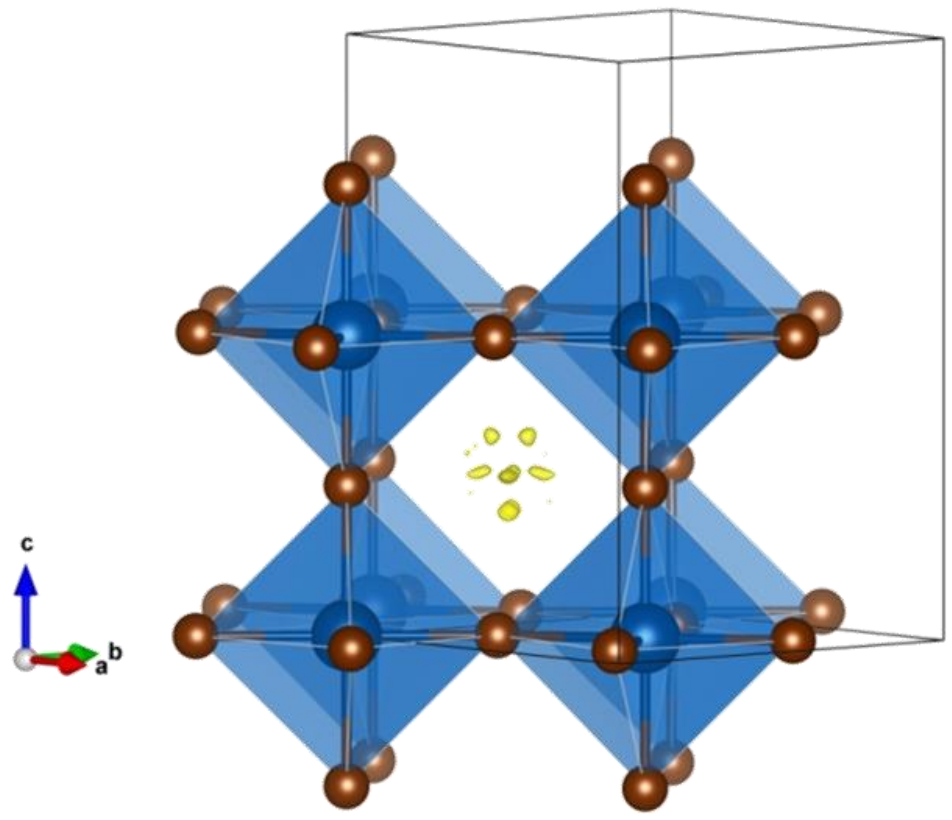

**Figure 3.** Schematic view of $MAPbBr_3$ at 210 K in the tetragonal system along [1-10]. The isosurface of electron density at 3.2 e is shown in yellow. The visualized atoms are limited in order to keep a view in accord to cubic system.

At 150 K some transient diffraction lines were observed, which could not be indexed either as tetragonal or as orthorhombic and must belong to an intermediate structure with a narrow stability range. At the lowest measured temperature, 120 K, the pattern was successfully refined in the orthorhombic symmetry (*Pnma*). The good agreement between observed and calculated profiles is shown in Figure S4, whereas the crystallographic parameters are listed in Table S6. As depicted in Figure 4, in this model the MA groups are not delocalized but oriented and contained in the (010) plane. Figures S5 and S6 additionally illustrate the structural evolution below RT.

To complement the SXRPD study, yielding information on the C and N positions of the MA units, a neutron diffraction investigation was essential to locate H atoms and elucidate the MA conformation in the lattice. Up to now, only three works are devoted to describe the methylammonium position (including the H atoms) in this perovskite. Two of them use single crystal neutron diffraction[24,25] and the third one uses the powder method, but with a deuterated sample.[23] In the present study, a NPD pattern was collected in a non-deuterated sample. Unquestionably, the fact that the incoherent scattering from hydrogen atoms generates a large background, together with a low Q contribution from the inelastic scattering of methyl groups might be inconvenient. Fortunately, this should not affect the quality of the Rietveld refinement if a correct statistic is reached for the crystallographic peaks.

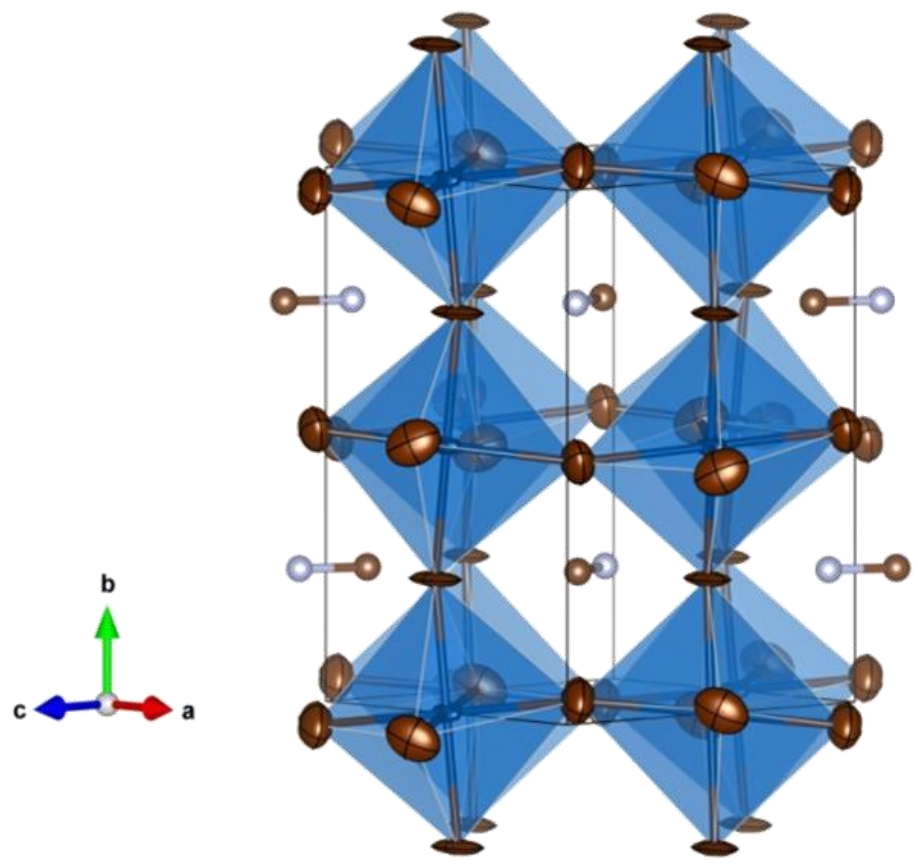

**Figure 4.** Schematic view of $MAPbBr_3$ at 120 K in the orthorhombic system. The visualized C and N atoms from MA are oriented within (101) plane. At this temperature the MA units are no longer delocalized.

At RT the initial model for the cubic structure concerned only the lead and bromine atoms, placed at 1a (0,0,0) and 3d (1/2,0,0) positions. This model gave a very poor fit to the neutron data, since the contribution of the H, C and N atoms is very important. At this point, a difference Fourier synthesis from the observed and calculated NPD data yielded the positive (yellow) and negative (blue) isosurfaces shown in Figure 5, which correspond to the H and C/N positions, respectively. These nuclear densities support that the MA units are delocalized in the A site of the perovskite. Considering the positive density, the C/N atoms are located at 24i (0.5,y,y) positions, as previously determined from SXRPD data. Then, the H positions can be elucidated from the geometric shape of methylammonium group. The observed geometry can be satisfied with two hydrogen atoms located at 24l (0.5,y,z) and 48n (x,y,z) Wyckoff sites. The Rietveld refinement of the positions of C/N and H atoms provides an excellent agreement between observed and calculated NPD data at RT; moreover, a marked improvement is obtained after the refinement of the displacement factors. Pb, Br and C/N atoms were refined with anisotropic displacement parameters while the H atoms with isotropic factors. The good fit between observed and calculated data is shown in Figure S7. The atomic parameters are listed in Table S7.

The obtained results agree with those previously informed by Mashiyama *et al.*[24] but subtly differ to that reported by Baikie *et al.*[25] The difference resides in the C/N positions. In the present work these atoms are located in 24*i* (0.5,y,y) site, while, in the Baikie *et al.* work, they were located in 24*l* (0.5,y,z) site. This may seem insignificant, but is very important to establish the orientations of methylammonium group in the lattice. The possible orientations were already studied and classified in models A, B or C where the MA are oriented along [100], [110] and [111] directions, respectively.[29] As can be seen, the preferred orientation in our case corresponds to the model B. In this case the MA distribution in the cubic phase can be deconvoluted in different directions at room temperature or at 240 K. Figure 6 represents the possible orientations in which MA resonates along [110] direction at room temperature. On the other hand, at 240 K the MA orientations are more restricted and respond to model A.[29] Figures S8 and S9 illustrate the different directions in which the MA can be found at RT and at 240 K, respectively.

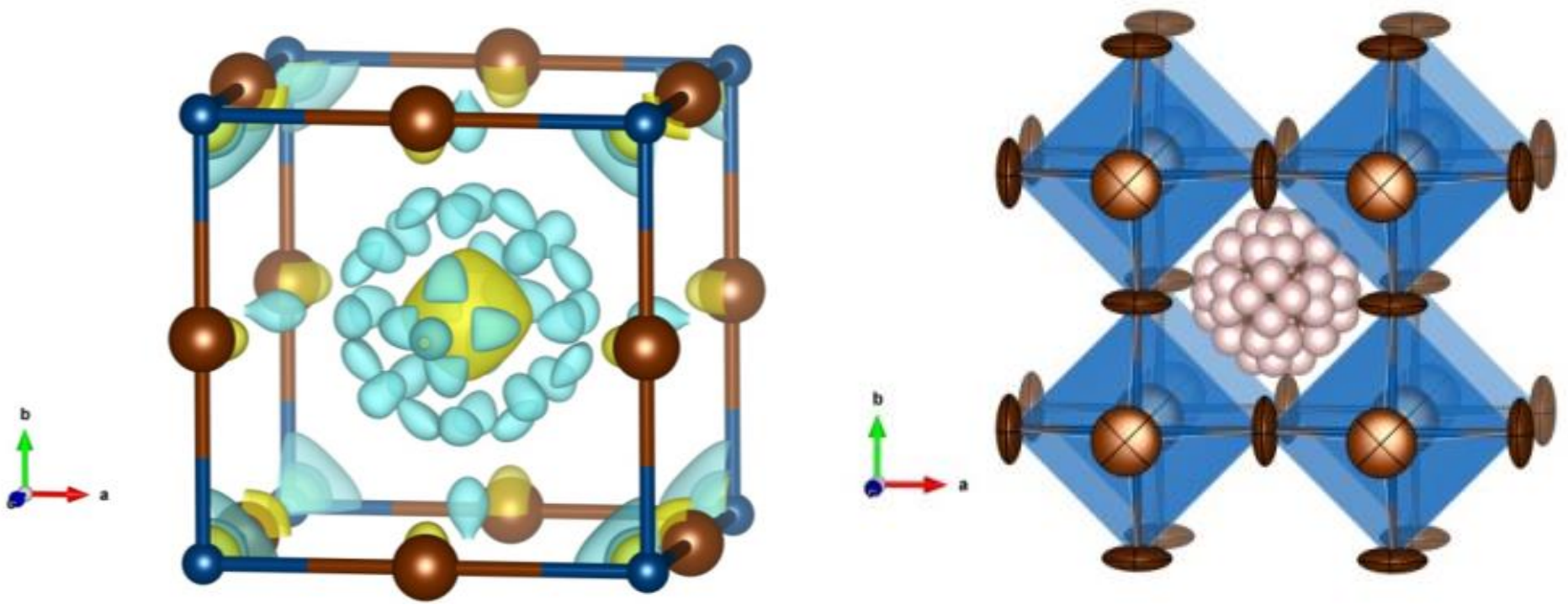

**Figure 5.** (left) Schematic view of nuclear density (scattering length density) of $MAPbBr_3$ at room temperature. The negative and positive isosurfaces are at 0.12 and 0.25 fm, respectively. (rigth) Crystal structure obtained from NPD refinement.

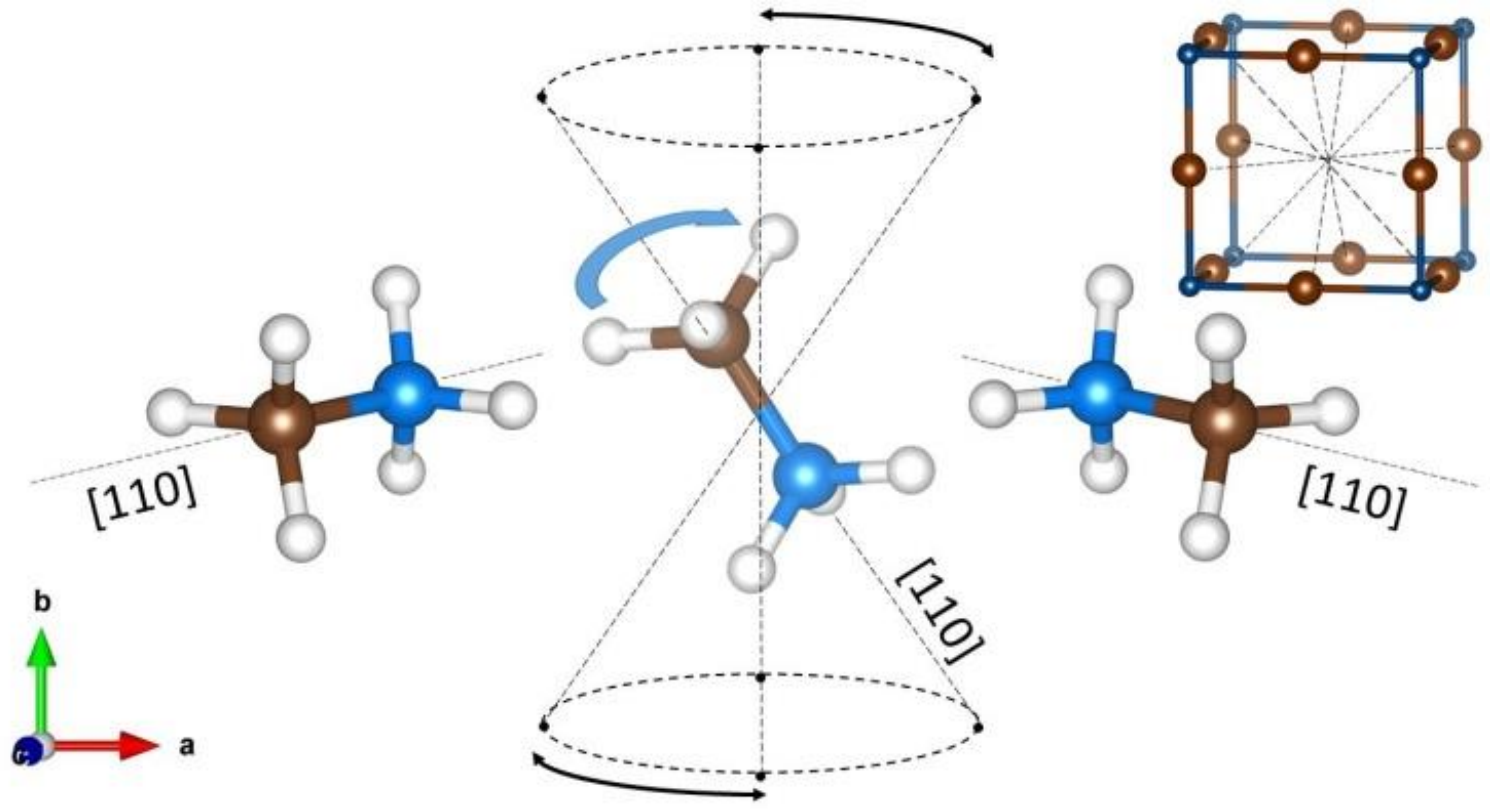

**Figure 6.** Schematic view of MA distribution in cubic phase at room temperature.

In order to illustrate the potential of the synthetized $CH_3NH_3PbBr_3$ in optoelectronic applications, we designed a photodetector device (Figure 7a) fabricated by drop-casting the perovskite solution onto Au/Cr pre-patterned electrodes. Figure 7b shows the current-voltage (IV) characteristics of the device under illumination at 505 nm for different incident optical power densities. The photoresponse exhibited by this device is among the largest ones reported for other perovkite-based photodetectors with different geometries: 2D-$MAPbI_3$,[30,31] thin films[32,33] nanowires[34,35] and networks.[36] We performed IVs measurements by employing light sources with different wavelengths all at the same illumination power of 1 mW/cm$^2$. We extract the wavelength responsivity spectrum of the photodetector (shown in Figure 7c), by using the next formula:

$$R = \frac{I_{ph}}{P}$$

being $R$ the responsivity, $I_{ph}$ the generated photocurrent and $P$ the illumination power on the device. In this Figure, we observe a maximum in the responsivity (0.26 A/W) for light with wavelength of 505 nm, which is near to the wavelength with maximum solar irradiance. Finally, in Figure 7d we plot the current through the device as a function of time with a fixed voltage under pulsed illumination. This illumination mode allows us to characterize the response time of the photodetector, in this device <100 ms (limited by the experimental setup). The limit of the response time obtained is comparable with values previously reported.[30,31,37,38]

## COCLUSIONS

In summary, we show that $MAPbBr_3$ is a promising photodetector material with an excellent responsivity for the more intense radiation of the solar spectrum, around 500 nm. We have deepened into the knowledge of the crystal structure and answered questions regarding the MA conformation inside the perovskite cage, which conspicuously evolve across the phase transitions that this material experiences below room temperature. A partial delocalization in the RT cubic phase, where C and N atoms adopt large multiplicity positions, evolve to a localization in the orthorhombic structure at 120 K, where MA units are oriented within (101) plane. A profound knowledge of the crystal structure details seems essential to establish structure-property correlations that may drive to further improvements.

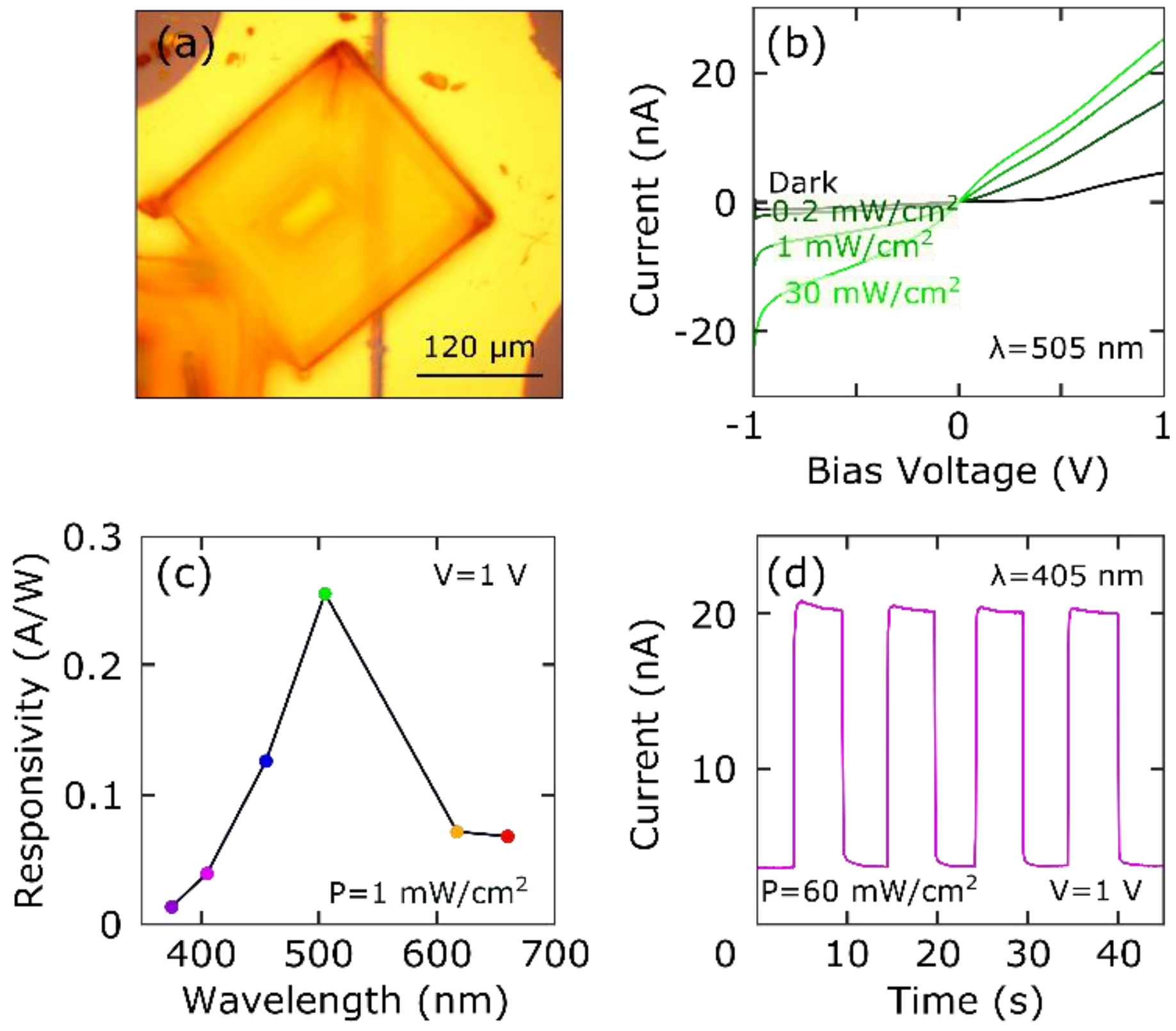

**Figure 7.** (a) Optical image of a photodetector device with a crystal grown between two pre-patterned electrodes. (b) Current-voltage (*IV*) curves under illumination at 505 nm for different incident optical power densities. (c) Responsivity spectrum for different wavelengths. (d) Photo current through the device as a function of time with a fixed voltage under modulated illumination.

## ACKNOWLEDGEMENTS

C.A. López acknowledges ANPCyT and UNSL for financial support (projects PICT2014-3576 and PROICO 2-2016), Argentine. C.A. López is a member of CONICET. J. A. Alonso thanks the Spanish MINECO for granting the project MAT2013-41099-R, Spain. PG and ACG acknowledge funding from the European Commission under the Graphene Flagship, contract CNECTICT-604391. The authors acknowledge ILL (France) and ALBA (Spain) for making all facilities available for the neutron and synchrotron diffraction experiments.